\documentclass[12pt,preprint]{aastex}

\newcommand{\likelihood}{\mathcal L}

\newcommand{\Mo}{$M_{\odot}$}

\newcommand{\MJ}{$M_{J}$}

\newcommand{\MAXLIMA}{{\footnotesize MAXLIMA}}

\shorttitle{Mass Distribution of Planets}
\shortauthors{Zucker \& Mazeh}

\begin{document}

\title{Derivation of the Mass Distribution of Extrasolar Planets
with \MAXLIMA\ - a Maximum Likelihood Algorithm}

\author{Shay Zucker and Tsevi Mazeh}
\affil{School of Physics and Astronomy, Raymond and Beverly Sackler
Faculty of Exact Sciences, Tel Aviv University, Tel Aviv, Israel}
\email{shay@wise.tau.ac.il; mazeh@wise7.tau.ac.il} 

\vskip 4.8 truecm

\begin{abstract}
We construct a maximum-likelihood algorithm - \MAXLIMA, to derive the
mass distribution of the extrasolar planets when only the minimum
masses are observed. The algorithm derives the distribution by solving
a numerically stable set of equations, and does {\it not} need any
iteration or smoothing. Based on 50 minimum masses, \MAXLIMA\ yields a
distribution which is approximately flat in $\log M$, and might rise
slightly towards lower masses. The frequency drops off very sharply
when going to masses higher than 10 \MJ, although we suspect there is
still a higher mass tail that extends up to probably 20 \MJ.  We
estimate that 5\% of the G stars in the solar neighborhood have
planets in the range of 1--10 \MJ\ with periods shorter than 1500
days.  For comparison we present the mass distribution of stellar
companions in the range of 100--1000\MJ, which is also approximately
flat in $\log M$. The two populations are separated by the
``brown-dwarf desert'', a fact that strongly supports the idea that
these are two distinct populations.  Accepting this definite
separation, we point out the conundrum concerning the similarities
between the period, eccentricity and even mass distribution of the two
populations.
\end{abstract}

\subjectheadings{binaries: spectroscopic --- 
methods: statistical ---
planetary systems ---
stars: low-mass, brown dwarfs ---
stars: statistics}

\section{INTRODUCTION}

Since the detection of the first few extrasolar planets their mass
distribution was recognized to be a key feature of the growing new
population. In particular, the potential of the high end of the mass
distribution to separate between planets on one side and brown dwarfs
and stellar companions on the other side was pointed out by numerous
studies (e.g., Basri \& Marcy 1997; Mayor, Queloz \& Udry 1998; Mazeh,
Goldberg \& Latham 1998).  A clear mass separation between the two
populations could even help to clarify one of the very basic questions
concerning the population of extrasolar planets --- the precise
definition of a planet (Burrows et al. 1997; see a detailed discussion
by Mazeh \& Zucker 2001). The lower end of the mass distribution could
indicate how many Saturn- and Neptune-like planets we expect on
the basis of the present discoveries, a still unsurveyed region of the
parameter space of extrasolar planets.

The present number of known extrasolar planets --- more than 60 are known
(Encyclopedia of extrasolar planets, Schneider 2001), offers an
opportunity to derive a better estimate of the mass distribution of this
population. In order to use the derived masses of the extrasolar planets
we have to correct for two effects. The first one is the unknown orbital
inclination, which renders the derived masses only {\it minimum}
masses. The second effect is due to the fact that stars with too small
radial-velocity amplitudes could not have been detected as
radial-velocity variables. Therefore, planets with masses too small,
orbital periods too large, or inclination angles too small are not
detected.

The effect of the unknown inclination of spectroscopic binaries was
studied by numerous papers (e.g., Mazeh \& Goldberg 1992; Heacox 1995;
Goldberg 2000), assuming random orientation in space. Heacox (1995)
calculates first the minimum mass distribution and then uses its
relation to the mass distribution to derive the latter. This
calculation amplifies the noise in the observed data, and necessitates
the use of smoothing to the observed data.  Mazeh \& Goldberg (1992)
introduce an iterative algorithm whose solution depends on the initial
guess.  In the present work we followed Tokovinin (1991, 1992) and
constructed an algorithm --- MAXimum LIkelihood MAss, to derive the
mass distribution of the extrasolar planets with a maximum likelihood
approach. \MAXLIMA\ assumes that the planes of motion of the planets
are randomly oriented in space and derives the mass distribution
directly by solving a set of numerically stable linear equations. It
does not require any smoothing of the data nor any iterative
algorithm. \MAXLIMA\ also offers a natural way to correct for the
undetected planets.

The randomness of the orbital planes of the discovered planets were
questioned recently by Han et al. (2001), based on the analysis of
Hipparcos data. However a few very recent studies (Pourbaix 2001;
Pourbaix \& Arenou 2001; Zucker \& Mazeh 2001a,b) showed that the
Hipparocos data do not prove the nonrandomness of the orbital
planes, allowing us to apply \MAXLIMA\ to the sample of known minimum
masses of the planet candidates.

In the course of preparing this paper for publication we have learned
about a similar paper by Jorissen, Mayor \& Udry (2001) that was
posted on the Astrophysics e-Print Archive (astro-ph).
Like Heacox (1995), Jorrisen et al. derive first the
distribution of the minimum masses and then apply two alternative algorithms to
invert it to the distribution of planet masses. One algorithm is a
formal solution of an Abel integral equation and the other is the
Richardson-Lucy algorithm (e.g., Heacox 1995). The first algorithm
necessitates some degree of data smoothing  and the second one 
requires a series of iterations.  The results of the first algorithm depend on
the degree of smoothing applied, and those of the second one on the
number of iterations performed.  \MAXLIMA\ has no built-in free parameter,
except the widths of the histogram bins. In addition, Jorissen et al.\ did not
apply any correction to the selection effect we consider here, and displayed
their results on a linear mass scale. We feel that a
logarithmic scale can illuminate some other aspects of the
distribution. Despite all the differences, our results are completely
consistent with those of Jorrisen et al., the sharp cutoff in the planet
mass distribution at about 10 \MJ, and the small high-mass tail that
extends up to about 20 \MJ\ in particular.

Section 2 presents \MAXLIMA, while section 3 presents our
results. Section 4 discusses briefly our findings.

\section{MAXLIMA}

\subsection{The Unknown Inclination}

Our goal is to estimate the probability density function (PDF) of the
secondary mass --- $f_m(m) dm$, given a set of observed minimum masses
\{$y_j; j=1,N\}$, where $y_j=m_j\times \sin i_j$, $m_j$ is the mass of
the j-th planet and $i_j$ is its inclination. Within certain
assumptions and limitations, \MAXLIMA\ finds the function $f_m$ that
maximizes the likelihood of observing these minimum masses.

Note that the present realization of \MAXLIMA\ uses the approximation that
the mass of the unseen companion is much smaller than the primary
mass. Within this approximation $y_j$ can be derived from the
observations of each system, given the primary mass.  In general, when
the secondary mass is not so small, the value of $m_j\times \sin i_j$
cannot be derived from the observations, and a more complicated
expression has to be used. Nevertheless, the extension of \MAXLIMA\ to
those cases is straightforward, and will be worked out in details in a
separate paper.

We assume that the directions of the angular momenta of the
systems are distributed isotropically, which will cause $\sin i$ to have
a PDF of the form:
$$
f_s(s)\,ds = \frac{s}{\sqrt{1-s^2}}\,ds \ ,
$$
where we denote $\sin i$ by $s$.
We further assume that the planet mass and its orbital inclination are
uncorrelated, and therefore the {\it joint} PDF of $\sin i$ and the
planet mass
have the form:
$$
f_{ms}(m,s)\,dm\,ds = f_m(m)\ \frac{s}{\sqrt{1-s^2}}\,dm\,ds \ .
$$

Now, define a variable $y=m\times \sin i =m\times s $, 
which has the PDF

$$
f_y(y)dy=
\int{ f_m(m){f_s(y/m)\over m}}\,dm\,dy = 
\int{ {f_m(m) \over m}{(y/m)\over \sqrt{1-(y/m)^2}} }\,dm\,dy \ .
$$
A sketch of contours of constant $y$ in the $m-s$ parameter space is
plotted in Figure~1.

We wish to estimate $f_m(m)$ in the form of a histogram with
K bins, between the limits  $m_{min} \leq m < m_{max}$.
We thus consider a partition of
the interval $[m_{min},m_{max})$:
$$
m_{min}=m_1,m_2,\ldots,m_{K+1}=m_{max}\ ,
$$
for which the PDF gets the form:
$$
f_m(m) \equiv b_k,\; 
{\rm if\ } m_k \leq m < m_{k+1},\; {\rm for\ } k=1,\ldots,K.
$$

Note, that the function $f_m(m)$ is supposed to be a probability density
function, and therefore its integral must equal one:
\begin{equation}
\sum_{k=1}^K b_k \Delta m_k = 1  \ ,
\end{equation}
where $\Delta m_k \equiv m_{k+1}-m_k$ for $k=1,\ldots,K$. 

The PDF of $y$ then gets the form:
\begin{eqnarray*}
f_y(y)\,dy &= &
\sum_{k=1}^K \int_{m_k}^{m_{k+1}} \frac{b_k}{m} 
\frac{y/m}{\sqrt{1-(y/m)^2}}\,dm\,dy \\
           &= &
\sum_{k=1}^K b_k \int_{m_k}^{m_{k+1}} \frac{1}{m} 
\frac{y/m}{\sqrt{1-(y/m)^2}}\,dm\,dy\ ,
\end{eqnarray*}
where the integrals do not depend on $f_m(m)$ at all, but only on the
intervals borders and $y$.

Now we can solve our problem in a maximum-likelihood fashion by
finding the set of $b_k$'s that maximizes the likelihood of the
actually observed values - ${y_j}$. The likelihood
function is:
\begin{eqnarray*} 
\likelihood(b_1,\ldots,b_K;y_1,\ldots,y_N) &= &\prod_{j=1}^N f_y(y_j) \\
&= &\prod_{j=1}^N \sum_{k=1}^K b_k \int_{m_k}^{m_{k+1}} \frac{1}{m} \frac{y_j/m}{\sqrt{1-(y_j/m)^2}}\,dm \\
&= &\prod_{j=1}^N \sum_{k=1}^K A_{jk} b_k \ ,
\end{eqnarray*}
where
\begin{equation}
A_{jk} = \cases {
 0                                        & $m_{k+1} \leq y_j$ \cr
 \int_{y_j}^{m_{k+1}} \frac{1}{m} 
\frac{y_j/m}{\sqrt{1-(y_j/m)^2}}\,dm      & $m_k \leq y_j < m_{k+1}$ \cr
\int_{m_k}^{m_{k+1}} \frac{1}{m} 
\frac{y_j/m}{\sqrt{1-(y_j/m)^2}}\,dm      & $y_j < m_k$ \ , \cr
}
\end{equation}
and the integrals are easily calculated even analytically.

In the appendix we present an elegant way to find the $b_k$'s that
maximize $\log \likelihood$ directly, without any iterations.

\subsection{Simulation}

In order to check the performance of \MAXLIMA\ and its realization we have
performed several simulations, some of which are presented in Figure
2. In those simulations we generated an artificial sample of planets
drawn from populations with different PDFs of the planet masses, and  
inclinations oriented isotropically in space. To make the simulation similar
to the present work we chose the size of each sample to be 50
planets. We assumed no selection effects. We then applied \MAXLIMA\ to the
simulated sample, the results of which are plotted in Figure~2.

The three examples of Figure~2 clearly show the power of MAXLIMA.

\subsection{Selection Effect: Undetected Planets}

We assume that the sample is constructed of planets with period, $P$,
between $P_{\rm min} \leq P \leq P_{\rm max}$. We further assume that the
search for planets discovered all radial-velocity variables with
amplitude $K$ larger than $K_{\rm min}$. We have to correct for planets
not detected because they induce $K$ smaller than the threshold. To do
that we note that the amplitude can be written as 
\begin{equation} 
K(P,M_p,M_1,\sin i) = 
204\,
\left( {P\over {{\rm day}}} \right) ^{-1/3}    \,  
\left( {M_1\over {M_{\odot}}} \right) ^{-2/3}    \, 
\left( {M_p\over {M_{Jup}}}   \right)            \, 
\sin i\ \,\   
\rm m\, s^{-1}                        \ . 
\end{equation} 

The expression $M_p\times \sin i$ is actually our $y$. For any {\it
given} value of $y$ and $M_1$ we can derive the maximum possibly
detected period --- $P_{\rm max-detect}$, given $K_{\rm min}$. This
implies that if we know the period distribution and we assume that the
period is uncorrelated to the mass distribution, we can estimate for each
of the given $y$'s the fraction of planets with long periods that were
not detected with the same $y$. This means that to correct for the
undetected planets with long periods we have to consider each of the j-th
detected systems as representing some $\alpha_j$ planets. If $P_{\rm
max-detect}$ is smaller than $P_{\rm max}$, then $\alpha_j$ is larger
than unity. Otherwise $\alpha_j$ is equal to unity.

We then can write a new generalized likelihood as:
\begin{eqnarray*} 
\likelihood(b_1,\ldots,b_K;y_1,\ldots,y_N;\alpha_1,\ldots,\alpha_N) 
&= &\prod_{j=1}^N \left( f_y(y_j) \right)^{\alpha_j}\\
&= &\prod_{j=1}^N \left( \sum_{k=1}^K b_k \int_{m_k}^{m_{k+1}} 
    \frac{1}{m} \frac{y_j/m}{\sqrt{1-(y_j/m)^2}}\,dm \right) ^{\alpha_j} \\
&= &\prod_{j=1}^N \left( \sum_{k=1}^K A_{jk} b_k \right)^{\alpha_j} \ .
\end{eqnarray*}
The Appendix shows an easy way used by \MAXLIMA\ to find a maximum to
$\log\likelihood$.

For simplicity we consider only circular orbits.  Eccentricity
introduces two effects (Mazeh, Latham \& Stefanik 1996), the first of
which is the dependence of $K$ on the eccentricity $e$.  Equation (3)
should include an additional factor of $(1- e^2)^{-1/2}$, which causes
$K$ to increase for increasing $e$.  The other factor is the dependence
of the detection threshold $K_{\rm min}$ on $e$.  Our simplifying
assumption about the constancy of $K_{\rm min}$ throughout the sample
breaks down when we consider eccentric orbits.  This is so because for
eccentric orbits the velocity variation tends to concentrate around the
periastron passage, and therefore $K_{\rm min}$ increases for increasing
eccentricity.  These two effects tend to cancel each other (Fischer \&
Marcy 1992), the net effect depends on the characteristics of the
observational search.  By running numerical simulations Mazeh et
al.\ (1996) have found that if the detection limit depends on the
r.m.s. scatter of the observed radial-velocity measurements, the two
effects cancel each other for any reasonable eccentricity. We therefore
chose not to include the eccentricity of the planets in our analysis.

\section{ANALYSIS AND RESULTS}

To apply \MAXLIMA\ to the current known sample of extrasolar planets
we considered all known planets and brown dwarfs as of April 2001. We
consider only G- or K-star primaries and therefore excluded Gls 876
from the sample.

Obviously, the present sample in not complete. In particular, not all
planets with long periods and small induced radial-velocity amplitudes
were discovered and/or announced. To acquire some degree of completeness
to our sample we have decided, somewhat arbitrarily, to exclude planets
with periods longer than 1500 days and with radial-velocity amplitudes
smaller than 40 m/s. The values of these two parameters determine
the correction of \MAXLIMA\ for the selection effect, for which we assumed
a period distribution which is flat in $\log P$. This choice of parameters
also implies that our analysis applies only to planets with periods
shorter than 1500 days. We further assumed that the primary mass is 1
\Mo\ for all systems.

In order to be consistent with the selection effects and the correction
we applied, we included in our analysis only planets that were discovered
by the high precision radial-velocity searches. We had to exclude
HD~114762 and similar objects that were discovered by other searches
(e.g., Latham et al.\ 1989; Mayor et al.\ 1992; Mazeh et al.\ 1996). 
This does not mean that we assume
anything about their nature in this stage of the study. Table~1 lists
all the known planets with G star primaries from the high precision
radial-velocity studies. Planets excluded from our analysis are marked
by an asterisk. We did not take into account the known inclination of
the planet around HD~209458. All together we are left with 50 planets.

The results of \MAXLIMA\ are presented in the lower panel of Figure~3 on a
logarithmic mass scale. Each bin is 0.3 dex wide, which means about a
factor 2 in mass. The value of each bin is the estimated number of
planets found in the corresponding range of masses in the known sample
of planets, after correcting for the undetected systems. To estimate the
uncertainty of each bin we ran 5000 Monte Carlo simulations and found
the r.m.s.\ of the derived values of each bin. Therefore, the errors
plotted in the figure represent only the statistical noise of the
sample. Obviously, any deviation from the assumptions of our model for
the selection effect induces further errors into the histogram, the
assumed period distribution in particular. This is specially true for
the first bin, where the actual number of systems is small and the
correction factor large. The value of the first bin is sensitive, for
example, to the assumed $K_{\rm min}$. Assuming $K_{\rm min}$ of 50 m/s
increased the value of the first bin by more than 50\%.

To compare the mass distribution of the planets with that of the stellar
secondaries we plot the latter on the same scale in an adjacent panel of
Figure~3. We plot here only two bins, with masses between 100 and 1000
\MJ. We follow the work of Mazeh (1999b) and Mazeh and Zucker (2001),
and used for those bins subsamples of binaries found by the CfA
radial-velocity search for spectroscopic binaries (Latham 1985) in the
Carney \& Latham (1987) sample of the high-proper-motion stars (Latham
et al.\ 2000; Goldberg et al.\ 2000). For the smaller bin we used a
subsample that included only the Galaxy disk stars (Goldberg 2000), and
for the larger-mass bin a subsample of this sample that included only
primaries with masses higher than 0.7 \Mo.  The values of those two bins
were derived with the algorithm of Mazeh \& Goldberg (1992).

Note that the upper panel {\it does not have any estimate of the values
of the bins with masses smaller than 100 \MJ}. This is so because the
CfA search does not have the sensitivity to detect secondaries in that
range. On the other hand, the lower panel does include information on
the bins below 100 \MJ. This panel presents the results of the
high-precision radial-velocity searches, and these searches could easily
detect stars with secondaries in the range of, say, 20--100 \MJ. We
assume that these binaries were not excluded from the various
radial-velocity searches at the first place, and further assume that
all, or at least most, findings of the various research groups
corresponding to this range of masses were already published. If these
two assumptions are correct, then the lower panel does represent the
frequency of secondaries in the mass range of 20--100 \MJ. This panel
shows that the frequency of secondaries in this range of masses is close
to zero. The present analysis is not able to tell whether this
``brown-dwarf desert'' extends up to 60, 80 or 100 \MJ.

The relative scaling of the planets and the stellar companions is not
well known. The spectroscopic binaries come from well defined samples --
577 stars for the lower-mass bin, and 312 stars for the higher-mass
stars (Goldberg 2000). However, this is not the case for the detected
planets, specially because the sample of published planets is not
complete and also because the search samples of the different groups are
not well documented in the public domain. We assumed, somewhat {\it arbitrarily},
that they come from a sample of 1000 stars, and scaled the stellar
bins accordingly. We also rescaled the stellar bins to account for the
fact that their bins are larger, and their period range extends up to 3000
days, assuming a flat distribution in $\log P$.  Therefore the values of
the stellar bins are our best estimate for the number of binaries for
1000 stars within a mass range of 0.3 dex, and up to a period of 1500
days.

Obviously the relative scaling of the two panels has a large
uncertainty. This scaling uncertainty is {\it not} reflected in the
error bars of the higher panel. Nevertheless we think that the
comparison is illuminating, as will be discussed in the next section.

\section{DISCUSSION}

The grand picture that is emerging from Figure~3 strongly indicates that
we have here two distinct populations. The two populations are separated
by a ``gap'' of about one decade of masses, in the range between 10 and
100 \MJ. Such a gap was already noticed by many early studies (Basri
\& Marcy 1997; Mayor, Queloz \& Udry 1998; Mayor, Udry \& Queloz 1998;
Marcy \& Butler 1998). Those early papers binned the mass distribution
linearly.  Here we follow our previous work (Mazeh et al.\ 1998)
and use a logarithmic scale to study the
mass distribution, because of the large range of masses, 0.5--1000 \MJ,
involved.  The logarithmic scale has also been used by Tokovinin (1992)
to study the secondary mass distribution in spectroscopic binaries, and
was suggested by Black (1998) to study the mass distribution of the
planetary-mass companions (see also Mazeh 1999a,b; Mazeh \& Zucker
2001; Mayor et al. 2001). 
The gap or the ``brown-dwarf desert'' are
consistent also with the finding of Halbwachs et al.\ (2000), who used
Hipparcos data and found that many of the known brown-dwarf candidates
are actually stellar companions.

We will assume that the two populations are the planets, at the low-mass
side of Figure~3, and the stellar companions at the high-mass end of the
figure.  Interestingly, the mass distribution of {\it single} stars
extends far below 100 \MJ\ (e.g., Zapatero Osorio et al.\ 2000; Lucas \& Roche
2000), indicating that the gap separating the two populations of {\it
companions} apparently does not exist in the population of single
stars/brown dwarfs. This difference probably indicates different
formation processes for single and secondary objects.

The distribution we derived in Figure~3 suggests that the planet mass
distribution is almost flat in $\log M$ over five bins --- from 0.3 to
10 \MJ. Actually, the figure suggests a slight rise of the distribution
towards smaller masses. The distinction between these two distributions
is not possible at this point, when our knowledge about planets with
sub-Jupiter masses is very limited. At the high-mass end of the planet
distribution the mass distribution dramatically drops off at 10 \MJ,
with a small high-end tail in the next bin. Although the results are
still consistent with zero, we feel that the small value beyond 10 \MJ\
might be real. The dramatic drop at 10 \MJ\ and the small high-mass tail
agree with the findings of Jorissen et al.\ (2001).

Examination of the two panels of Figure~3 suggests that per equal dex
range of masses the frequency of stellar secondaries is higher than that
of the planets by a factor of about 2. As we emphasized in the previous
section, this is a very preliminary result that should be checked by
future observations. Nevertheless, the frequency of planets is
impressive by itself. Our results indicate that about 5\% of the stars
have planets with masses between 1 and 10 \MJ. This is so because the
number of multiple planets in this sample is small, so the number of
planets considered in the figure is about the number of stars found to
have one or more planets. If this frequency extends further down the
mass axis to Earth masses, we might find that more than 10\% of the
stars have planets with periods shorter than 1500 days.

The analysis presented here raises the question what mechanism can
produce flat or approximately flat mass distribution of planets up to 10
\MJ.  What determines the mass of the forming planet? The
present paradigm assumes that planets were formed out of a
protoplanetary disk. Is it the mass, density, angular momentum or the
viscosity of the disk that determined the planet mass? If planets were
formed by accreting gas onto a rocky core, is the planet mass determined
also by the location or evolutionary phase of the formation of the rocky
core?  Any detailed model of planet formation should account for this
mass distribution.

The clear distinction between the two populations suggests that planets
and stellar companions were formed by two different processes.  This is
so despite the striking similarity between the distributions of the
eccentricities and periods of the two populations (Heacox 1999;
Stepinski \& Black 2000, 2001; Mayor \& Udry 2000; Mazeh \& Zucker
2001). Even the mass distributions of the two populations might be very
similar --- approximately flat in $\log M$. This is still a conundrum that
any formation model for planets as well as for binaries needs to solve.

\acknowledgments

We are indebted to Yoav Benjamini for illuminating discussions.  This
work was supported by the US-Israel Binational Science Foundation
through grant 97-00460 and the Israeli Science Foundation (grant
no. 40/00)

\appendix
\section*{Appendix}

We want to find the maximum likelihood to observe a given set of
observed minimum masses $\{y_j;j=1,N\}$. As usual, it is easier to
maximize the logarithm of the likelihood function:
$$
\log \likelihood = \sum_{j=1}^N \alpha_j \times 
                   \log \left( \sum_{k=1}^K A_{jk} b_k \right) \ ,
$$
where each $A_{jk}$ depends on the corresponding $y_j$ through
Equation (2).

The $b_k$'s are not all independent. They are constrained by Equation
(1), and therefore we modify
our target function by adding a Lagrange multiplier term:
$$
\log \likelihood = \sum_{j=1}^N  \alpha_j \times
                   \log \left( \sum_{k=1}^K A_{jk} b_k \right) + 
    \lambda \left( \sum_{k=1}^K b_k \Delta m_k - 1 \right) \ .
$$

The optimization is performed by equating the partial derivatives of
this target function to zero:
\begin{eqnarray*}
 \frac{\partial \log \likelihood}{\partial b_k} &= 
&\sum_{j=1}^N
\frac{  \alpha_j \times A_{jk}}{\sum_{l=1}^K A_{jl} b_l} + 
                                \lambda \Delta m_k = 0, \ 
{\rm for\ } k=1,\ldots,K \\
 \frac{\partial \log \likelihood}{\partial \lambda} &= 
&\sum_{k=1}^K b_k \Delta m_k -1 = 0 \ .
\end{eqnarray*}
The parameter 
$\lambda$ is eliminated quite easily. We first multiply each of the
$K$ equations by the corresponding $b_k$ and then sum them up
to get:
$$
\sum_{k=1}^K \sum_{j=1}^N\frac{   \alpha_j \times A_{jk} b_k}
          {\sum_{l=1}^K A_{jl} b_l} +
  \lambda \sum_{k=1}^K b_k \Delta m_k = 0 \ .
$$
Changing the order of summation reduces the first term, after a simple
manipulation, to simply 
$$
N_{eff}=\sum_{j=1}^N \alpha_j \ .
$$
Using the constraint reduces the second
term to $\lambda$, and we finally get:
$$
N_{eff} + \lambda = 0 \ ,
$$
and we can simply set $\lambda = -N_{eff}$. The $K$ equations we are now left
with are:
$$
\sum_{j=1}^N  \alpha_j \times A_{jk} 
\frac{1}{\sum_{l=1}^K A_{jl} b_l} = N_{eff} \Delta m_k \ .
$$
We have a set of $K$ non-linear equations in $K$ variables - $b_k$'s. 

An elegant
reduction of the complexity of the problem can be achieved if we set $K$
to $N$, by assigning $m_k \equiv y_k$ for $k = 1,\ldots,N$. Let us
also denote (remember that now $K=N$):
\begin{equation}
\label{E3} h_j \equiv \frac{1}{\sum_{l=1}^N A_{jl} b_l}
\end{equation}
The $N$ equations now look like:
\begin{equation}
\label{E4} \sum_{j=1}^N  \alpha_j \times A_{jk} h_j = N_{eff} \Delta m_k \ ,
\end{equation}
which is a system of $N$ linear equations in the $N$ variables
$h_j$. We can easily solve for them. The problem is even more 
easily solved when we note that the matrix $A_{jk}$ is upper-triangular
and thus the amount of computations needed for the solution is
smaller. Furthermore, examination of the integrals involved in
the calculation of
$A_{jk}$ (Eq. 2) shows that the matrix is very close to being 
diagonally-dominated, and therefore the set of linear equations is
numerically stable. Having solved for $h_j$ we face again a similar system of
linear equations in order to solve for $b_k$, coming from the
definition of $h_j$:
$$
\sum_{l=1}^N A_{jl} b_l = \frac{1}{h_j} .
$$
In this system the matrix is the transposed matrix of the
previous system of linear equations if the $\alpha_j$'s are all
equal to unity.

Obviously, we wish to estimate the densities of our original
intervals. But these are easily calculable from the $N$ densities $b_k$
calculated above, as simple linear combinations.

\begin{figure}
\plotone{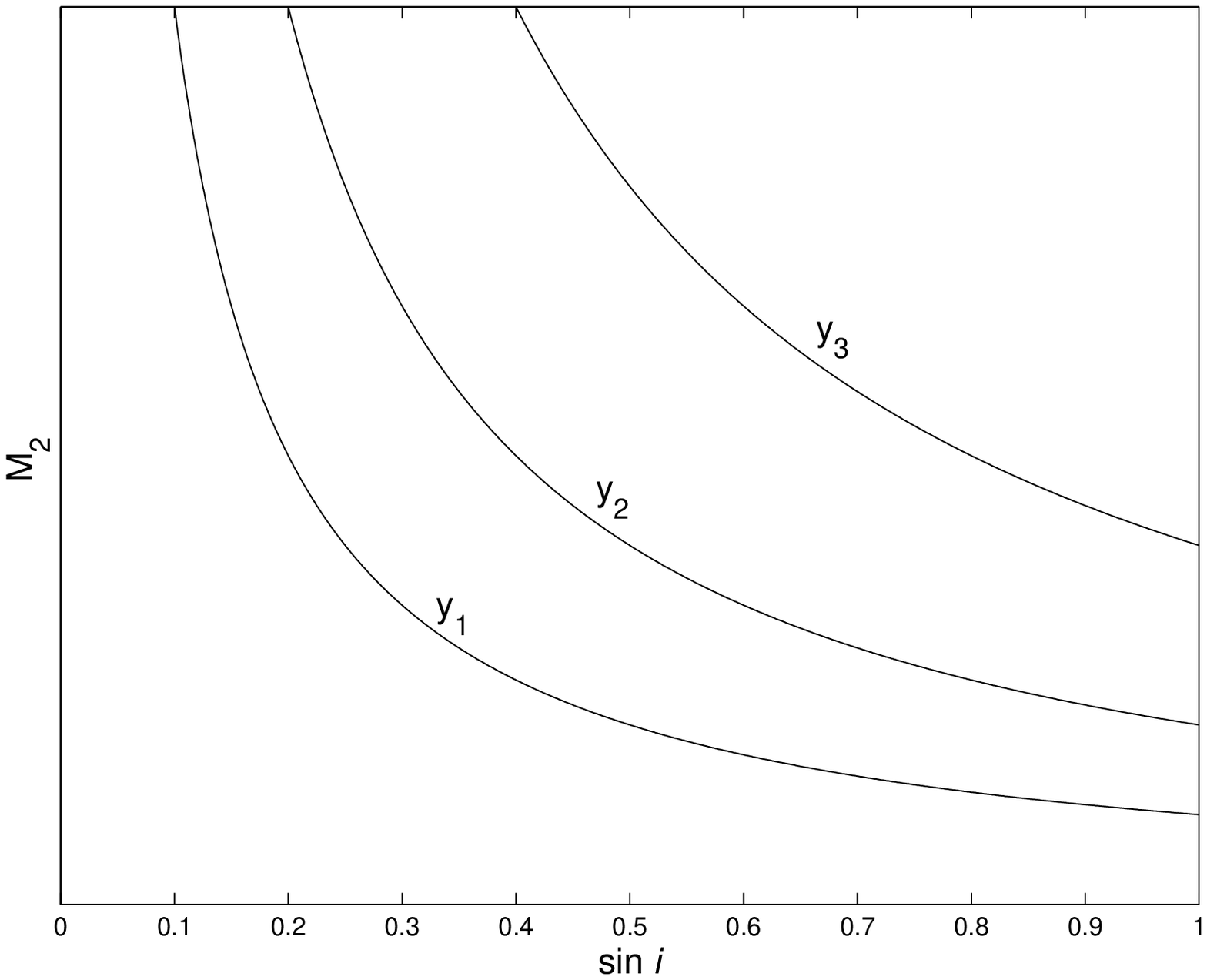}
\caption{Three contours of constant $y$'s in the $m-s$
plane.}
\end{figure}

\begin{figure}
\epsscale{0.35}
\plotone{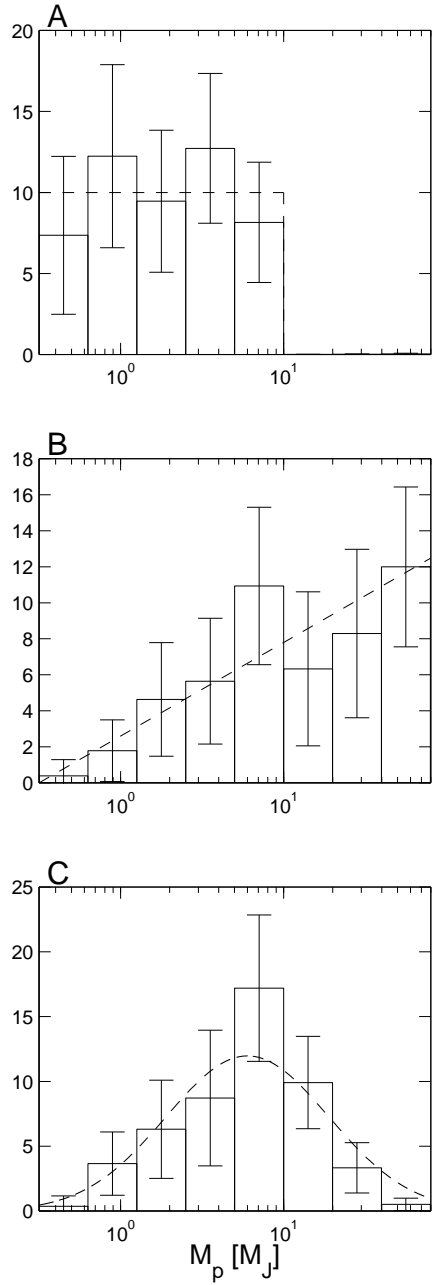}
\caption{Three simulations of MAXLIMA. In all three panels the dashed
line shows the input distribution and the histogram the results of
MAXLIMA.}
\end{figure}

\begin{figure}
\epsscale{1.0}
\plotone{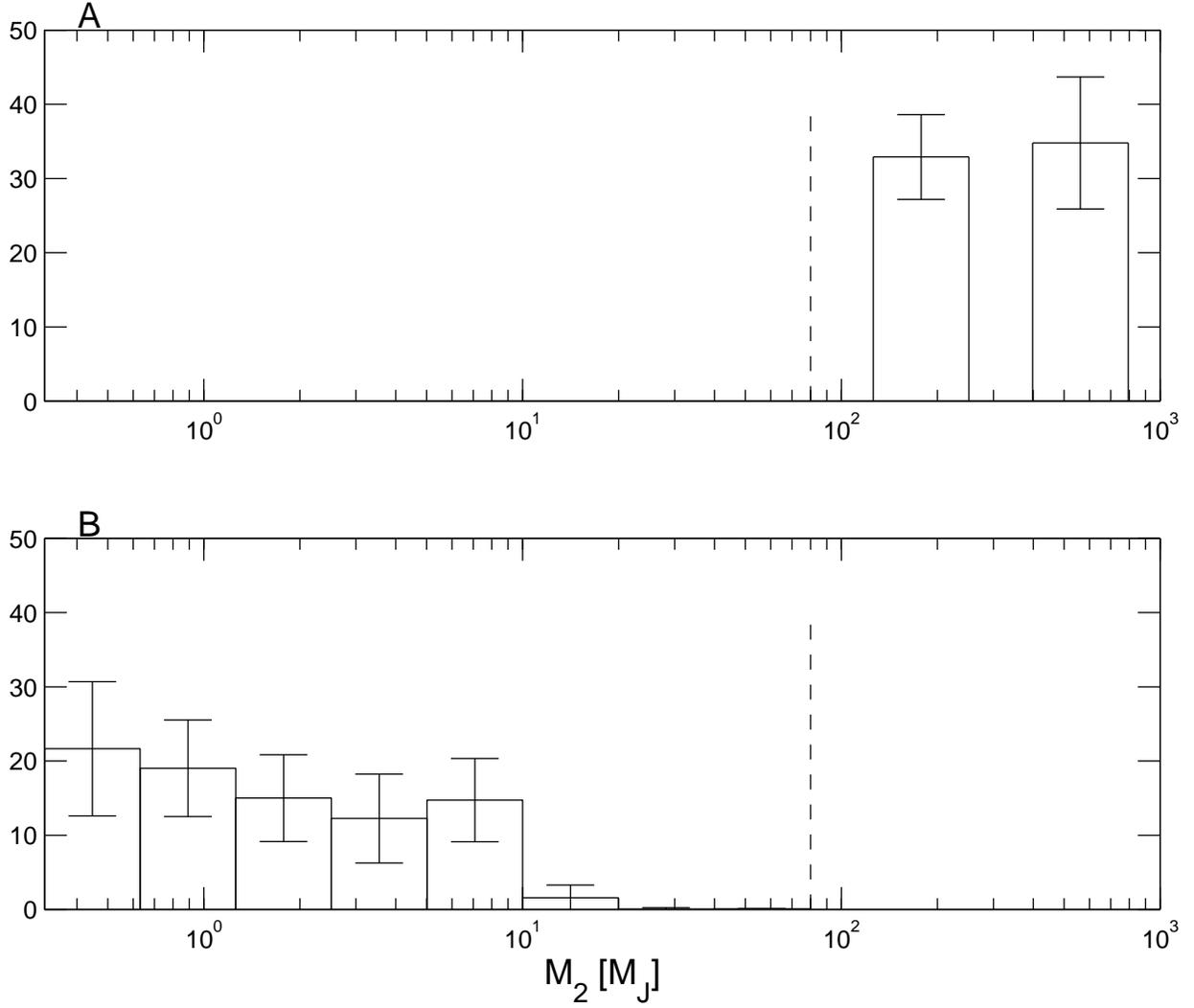}
\caption{The mass distributions of the planets and the stellar
companions}
\end{figure}

\begin{deluxetable}{llcc}
\tabletypesize{\scriptsize}
\tablewidth{0pt}
\tablecolumns{4}
\tablecaption{Substellar companions list}
\tablehead{
\colhead{Name} & 
\colhead{$M\sin i$} & 
\colhead{$P$} &
\colhead{$K$} \\
\colhead{} &
\colhead{(\MJ)} &
\colhead{(days)} &
\colhead{(m\ s$^{-1}$)}
}
\startdata
$^\ast$HD 83443 b     & 0.16  & 29.83  & 14 \\
$^\ast$HD 16141       & 0.215 & 75.82  & 11 \\
$^\ast$HD 168746      & 0.24  & 6.41   & 28 \\
$^\ast$HD 46375       & 0.249 & 3.02   & 35 \\
HD 83443 c     & 0.34  & 2.985  & 56 \\
$^\ast$HD 108147      & 0.34  & 10.88  & 37 \\
HD 75289       & 0.42  & 3.51   & 54 \\
51 Peg         & 0.47  & 4.23   & 56 \\
BD -10\degr 3166& 0.48 & 3.487  & 61 \\
$^\ast$HD 6434        & 0.48  & 22.09  & 37 \\
HD 187123      & 0.48  & 3.10   & 69 \\
$^\ast$Gliese 876 c   & 0.56  & 30.12  & 81 \\
HD 209458      & 0.69  & 3.52   & 86 \\
$\upsilon$ And b&0.69  & 4.617  & 71 \\
HD 192263      & 0.787 & 24.36  & 68 \\
HD 38529       & 0.81  & 14.32  & 54 \\
HD 179949      & 0.84  & 3.09   & 101 \\
55 Cnc         & 0.84  & 14.65  & 77 \\
$^\ast\epsilon$ Eri & 0.86  & 2502.1 & 19 \\
$^\ast$HD 82943 c     & 0.88  & 222    & 34 \\
HD 121504      & 0.89  & 64.6   & 45 \\
HD 130322      & 1.02  & 10.72  & 115 \\
HD 37124       & 1.04  & 155.7  & 43 \\
$\rho$ CrB     & 1.1   & 39.65  & 67 \\
HD 52265       & 1.13  & 118.96 & 45 \\
$^\ast$HD 177830      & 1.22  & 391.6  & 34 \\
HD 217107      & 1.27  & 7.13   & 140 \\
HD 210277      & 1.28  & 437    & 41 \\
$^\ast$HD 27442       & 1.43  & 426.5  & 34 \\
16 Cyg B       & 1.5   & 801    & 44 \\
HD 74156 b     & 1.56  & 51.6   & 108 \\
HD 134987      & 1.58  & 259.6  & 50 \\
HD 82943 b     & 1.63  & 445    & 46 \\
$^\ast$Gliese 876 b   & 1.89  & 61.02  & 210 \\
HD 160691      & 1.97  & 743    & 54 \\
HD 19994       & 2.0   & 454    & 45 \\
HD 213240      & 3.7   & 759    & 91 \\
$\upsilon$ And c&2.06  & 240.6  & 58 \\
HD 8574        & 2.23  & 228.8  & 76 \\
HR 810         & 2.26  & 320.1  & 67 \\
47 UMa         & 2.39  & 1090   & 45 \\
HD 12661       & 2.79  & 252.7  & 88 \\
HD 169830      & 2.96  & 230.4  & 83 \\
$^\ast$14 Her         & 3.3   & 1654    & 73 \\
GJ 3021        & 3.32  & 133.82 & 164 \\
HD 92788       & 3.34  & 326.7  & 100 \\
HD 80606       & 3.41  & 111.8  & 414 \\
HD 195019      & 3.47  & 18.2   & 272 \\
$\tau$ Boo     & 3.87  & 3.31   & 469 \\
Gliese 86      & 4     & 15.78  & 380 \\
$\upsilon$ And d& 4.10 & 1313   & 68 \\
HD 50554       & 4.9   & 1279   & 95 \\
HD 190228      & 5     & 1161   & 95 \\
HD 222582      & 5.3   & 575.9  & 184 \\
HD 28185       & 5.6   & 385    & 168 \\
HD 10697       & 6.35  & 1072.3 & 119 \\
HD 178911 B    & 6.46  & 71.50  & 343 \\
70 Vir         & 6.6   & 116.7  & 318 \\
HD 106252      & 6.81  & 1500   & 139  \\
HD 89744       & 7.2   & 256    & 257 \\
HD 168443 b    & 7.2   & 58.12  & 473 \\
$^\ast$HD 74156 c     & $>$7.5& 2300   & 121 \\
HD 141937      & 9.7   & 659    & 247 \\
$^\ast$HD 114762      & 11    & 84.03  & 600 \\
HD 202206      & 14.7  & 259    & 554 \\
$^\ast$HD 168443 c    & 15.1  & 1667   & 288 \\
$^\ast$HD 127506      & 36    & 2599   & 891 \\
\enddata
\end{deluxetable}

\end{document}